\documentclass[aps,prb,reprint,amssymb,amsmath,superscriptaddress]{revtex4-1}

\usepackage{graphicx}
\usepackage[english]{babel}
\bibpunct{[}{]}{,}{n}{}{}

\begin{document}

\title{The valence band energy spectrum of HgTe quantum wells with inverted band structures}

\author{G.\,M.~Minkov}
\affiliation{Institute of Natural Sciences and Mathematics, Ural Federal University,
620002 Ekaterinburg, Russia}

\author{V.\,Ya.~Aleshkin}
\affiliation{Institute for Physics of Microstructures  RAS, Nizhny Novgorod, Russia}

\author{O.\,E.~Rut}
\affiliation{Institute of Natural Sciences and Mathematics, Ural Federal University,
620002 Ekaterinburg, Russia}

\author{A.\,A.~Sherstobitov}

\affiliation{Institute of Natural Sciences and Mathematics, Ural Federal University,
620002 Ekaterinburg, Russia}

\author{A.\,V.~Germanenko}

\affiliation{Institute of Natural Sciences and Mathematics, Ural Federal University,
620002 Ekaterinburg, Russia}

\author{S.\,A.~Dvoretski}

\affiliation{Institute of Semiconductor Physics RAS, 630090
Novosibirsk, Russia}

\author{N.\,N.~Mikhailov}

\affiliation{Institute of Semiconductor Physics RAS, 630090
Novosibirsk, Russia}
\affiliation{Novosibirsk State University, Novosibirsk 630090, Russia}

\date{\today}

\begin{abstract}
The energy spectrum of the valence band in HgTe/Cd$_x$Hg$_{1-x}$Te quantum wells with a width $(8-20)$~nm has been studied experimentally by magnetotransport effects and  theoretically in framework $4$-bands $kP$-method. Comparison of the Hall density with the density found from period of the Shubnikov-de Haas (SdH) oscillations clearly shows that the degeneracy of states  of the top of the valence band is equal to 2 at the hole  density  $p< 5.5\times 10^{11}$~cm$^{-2}$. Such degeneracy does not agree with the calculations of the spectrum performed within the framework of the $4$-bands $kP$-method for symmetric quantum wells. These calculations show that the top of the valence band consists of four spin-degenerate  extremes located at $k\neq 0$ (valleys) which gives the total degeneracy $K=8$.  It is shown that taking into account the ``mixing of states'' at the interfaces  leads to the removal of the spin degeneracy that reduces the degeneracy to $K=4$. Accounting for any additional asymmetry, for example, due to  the difference in the mixing parameters at the interfaces, the different broadening of the boundaries of the well, etc, leads to reduction of the valleys degeneracy, making $K=2$. It is noteworthy that for our case two-fold degeneracy occurs due to degeneracy of two single-spin valleys.
The hole effective mass  ($m_h$)  determined from analysis of the temperature dependence of the amplitude of the SdH oscillations show that $m_h$ is equal to $(0.25\pm0.02)\,m_0$ and weakly increases with the hole density. Such a value of $m_h$ and its dependence on the hole density are in a good agreement with the calculated effective mass.
\end{abstract}


\maketitle

\section{Introduction}
\label{sec:intr}

The quantum wells in heterostructures HgTe/Cd$_x$Hg$_{1-x}$Te have a number of unusual properties compared to quantum wells based on  semiconductors with non-zero band gap. The reason for this is the ``negative'' band gap of gapless semiconductor HgTe  in which $\Gamma_6$ band, forming a conduction band in conventional  semiconductors, is located below the $\Gamma_8$ band, which in conventional semiconductors forms the valence band. Great attention is now paid to the study of topological states arising in such wells (see recent paper \cite{TopIns}).	Knowledge of the energy spectrum of two-dimensional (2D) carriers is required for reliable interpretation of all phenomena and it has been studied in the last decade, both theoretically \cite{Gerchikov90,Zhang01,Novik05,Bernevig06,ZholudevPhD,Tarasenko15} and experimentally \cite{Landwehr00,Zhang02,Ortner02,Zhang04,Koenig07,Gusev11,Kvon11}. However, quite a lot of differences between the experimental data and the results of the calculations has been accumulated  to date. First of all it refers to the spectrum of the top of valence band in the structures with quantum well width $d>(7.5-8)$~nm \cite{Kozlov11,Kvon11-1,Minkov13,Olshanetsky12}. The multi-band \emph{kP}-calculations predict for such a case the nonmonotone $E$~versus~$k$ dependence and the existence of four local maxima (valleys) at $k\neq0$ and so the eight-fold degeneracy of these states (two-fold spin degeneracy, $s=2$, and four-fold valley degeneracy,  $v=4$). However, such a degeneracy was not justified experimentally.

\begin{figure}
\includegraphics[width=0.8\linewidth,clip=true]{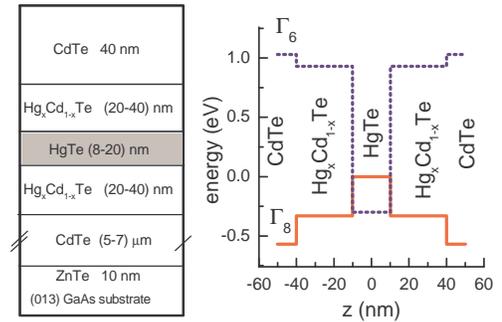}
\caption{(Color online) Sketch and energy diagram of the structures investigated. }
\label{F1}
\end{figure}

So,  it should be recognized that the band spectrum of these structures cannot be considered well established at the moment despite the large number of papers devoted to the study of HgTe/Cd$_x$Hg$_{1-x}$Te quantum well heterostructures. In this paper, we present the results of experimental study of the magnetotransport in the heterostructures with the HgTe quantum wells with $d$ from $8.3$~nm to $20.4$~nm, within a wide hole density range. The comparison of the Hall density and density found from the SdH  oscillations  clearly shows that the states near the top of valence band are two-fold degenerate within whole studied range of hole density and quantum well width. It is shown that a hole effective mass ($m_h$) found from the temperature dependence of the amplitude of SdH oscillations is about $(0.25\pm0.02)\,m_0$ at density $(1-2)\times 10^{11}$~cm$^{-2}$ and slightly increases with the increasing hole density ($m_0$ is the free electron mass).
To understand and interpret the experimental data we have calculated the energy spectrum using a $4$-band Kane Hamiltonian. We show that taking into account the interface inversion asymmetry and any additional asymmetry, for example, due to  the difference in the mixing parameters at the interfaces, the different broadening of the boundaries of the well, etc., explains the two-fold degeneracy of the states and gives quantitative agreement with the value of the hole effective mass and its density dependence.

\section{Experiment}
\label{sec:expdet}

Our HgTe quantum wells were realized on the basis of HgTe/Cd$_{x}$Hg$_{1-x}$Te ($x=0.5-0.7$)  heterostructure grown by molecular beam epitaxy on GaAs substrate with the ($013$) surface orientation \cite{Mikhailov06}. The nominal widths of the quantum wells under study were $d = (8.3-20.4)$~nm. The samples were mesa etched into standard Hall bars of $0.5$~mm width with the distance between the potential probes of $0.5$~mm. To change and control the electron and hole densities  in the quantum well ($n$ and $p$, respectively), the field-effect transistors were fabricated with parylene as an insulator and aluminium as a gate electrode. The measurements were performed at
temperature  $(1.3 - 4.2)$~K in the  magnetic field up to $6$~T. For each heterostructure, several samples were fabricated and studied. The sketch and the energy diagram of the structure investigated are shown in Fig.~\ref{F1}.

The  parameters of the structures investigated are presented in the Table~\ref{tab1}.
The main results for all the structures investigated are close to each other, therefore, as an example, let  us consider in more detail the data obtained for the structure H1529  with $d=8.3$~nm.

\begin{figure}
\includegraphics[width=\linewidth,clip=true]{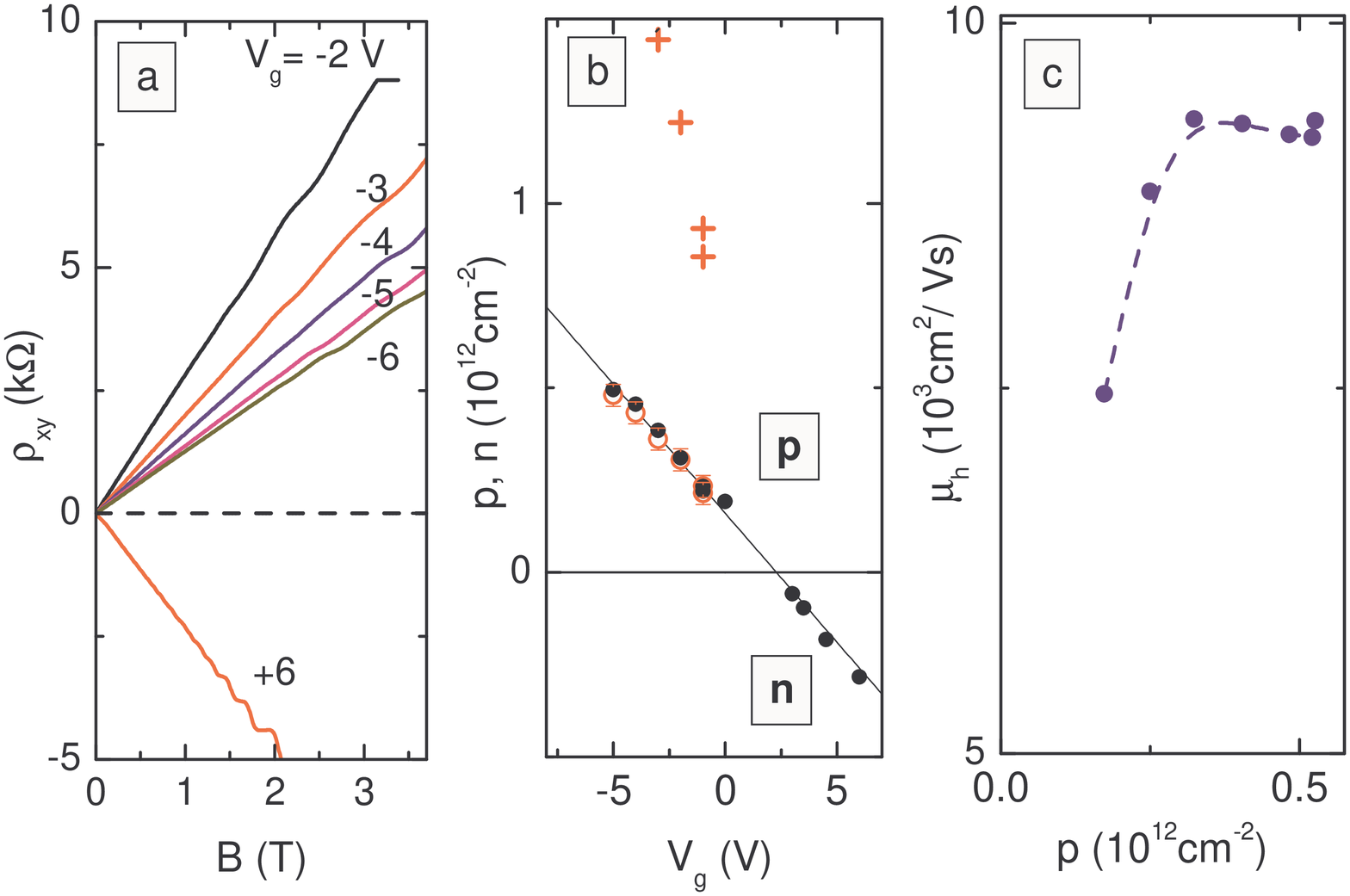}
\caption{(Color online) (a) -- The magnetic field dependencies of  $\rho_{xy}$ for some $V_g$ values. (b) -- The gate voltage dependence of the carriers density. The solid circles are the Hall density for holes $p_H=1/eR_H(B=1\text{~T})$ and for electrons $n_H=-1/eR_H(B=0.2\text{~T})$.  The crosses and empty circles are the hole density found from SdH oscillations for the degeneracy of the Landau levels (spin and valleys) 8 and 2 (crosses and circles, respectively).  The solid line is carrier density found from capacitance between the 2D gas and gate electrode as $C(V_g^\text{CNP}-V_g)/eS_g$, where $V_g^\text{CNP}$ is the gate voltage corresponding to charge neutrality point.  (c) -- The density dependence of the hole mobility. $T=1.4$~K.}
\label{F2}
\end{figure}

First, let us consider the behavior of the transverse ($\rho_{xy}$) and longitudinal  ($\rho_{xx}$) resistivity within classical magnetic fields at different gate voltages ($V_g$). As seen from
Fig.~\ref{F2}{(a)},  $\rho_{xy}(B)$ changes the sign when the gate voltage changes. The value of $\rho_{xy}$ linearly depends on magnetic field in $B\lesssim 1$~T within hole conductivity range (when $V_g <-1$~V)  and in  $B< (0.4-0.5)$~T within electron conductivity range  (that corresponds to $V_g> + 2$~V). (An exception is the structures with $d=15$~nm and $20.4$~nm where  two-types conductivity is observed in magnetic field $B<0.1$~T due to overlapping of the conduction and valence bands.) The gate voltage dependencies of the Hall density of the holes, $p_H=1/eR_H(1\text{~T})$, and electron density, $n_H=-1/eR_H(0.2\text{~T})$, plotted in Fig.~\ref{F2}(b) show that the rate of change of the density of the electrons and holes, $-dn_H/dV_g$ and $dp_H/dV_g$, respectively, are close to each other and they are well described by the dependence  $(1.6\times 10^{11}-7\times 10^{10} V_g, \text{\,V})$,~cm$^{-2}$ over the whole gate voltage range. It should be noted that the measurements of the capacitance ($C$) between the gate electrode and two-dimensional gas in the quantum well  for the same sample give the  value of $(1/e)dQ/dV_g=C/S_g=(7\pm 0.1)\times10^{10}$~cm$^{-2}$/V (where $S_g$ is the gate area). This value within experimental accuracy coincides with  $dn/dV_g$ and $-dp/dV_g$ found from the Hall measurements (the quantum capacitance gives contribution less than $1.5$ percent).

\begin{table}
\caption{The parameters of  heterostructures under study}
\label{tab1}
\begin{ruledtabular}
\begin{tabular}{ccccc}
number & structure & $d$ (nm) & type & \\
\colrule
  1& H1529 & 8.3   & $p$     \\
  2& H725 & 8.3   & $p, n^a$     \\
  3& HT71 & 9.5     & $p, n^a$    \\
  4& H1524 & 10     & $n$    \\
  5& H1312 & 15     & $p$     \\
  6& H1114 & 20.4     & $p$     \\
\end{tabular}
\end{ruledtabular}
\footnotetext[1]{After illumination.}
\end{table}

So, $1/e|R_H|$ gives the density of the holes and electrons.
The values of the hole mobility are large enough, $\mu_h=(7.5-8.5)\times 10^3$~cm$^2$/V\,s [see Fig.~\ref{F2}(c)] and as Fig.~\ref{F3}(a) shows the SdH oscillations are observed at
$T=1.4$~K. It is seen that  some distortion of the oscillations starts to be observed in magnetic fields higher than $\simeq 2.5$~T. We
assume, that  this is due to the complex spectrum of the valence band in magnetic field, therefore we restrict  ourselves for further consideration to the range of relatively low  magnetic fields   $B<2.5$~T. In Fig.~\ref{F3}(b), we have plotted the positions of the maxima of $\rho_{xx}$ at different $V_g$ values. One can see that the maxima shift to the higher fields with the growing hole density  when $V_g$ becomes more negative. The data are well extrapolated  to  $V_g\simeq 2.3$~V at  $B=0$, i.e., to the $V_g$ value, which corresponds to the charge neutrality point [compare Fig.~\ref{F2}(b) and Fig.~\ref{F3}(b)].

Now let us find the hole density from the SdH oscillations. For the case when the number of undistorted oscillations is not large, it is more reliable to determine their frequency in the reciprocal magnetic field from the fitting of the experimental curves to the Lifshits-Kosevich formula \cite{LifKos55} rather than from the Fourier spectrum:
\begin{equation}
\frac{\rho_{xx}(B,T)-\rho_{xx}(0,T)}{\rho_{xx}(0,T)}\propto A(B,T)\,\cos{\left(\frac{2\pi f}{B}\right)}
\label{eq01}
\end{equation}
with
\begin{equation}
A(B,T)=\exp\left( -\frac{\Delta}{\hbar \omega_c}\right)\, \frac{2\pi ^2 k_BT}{\hbar \omega_c}{\Big \slash}\sinh\left(\frac{2\pi ^2 k_BT}{\hbar \omega_c}\right),
\label{eq02}
\end{equation}
where $f$ is the oscillations frequency in the reciprocal magnetic field,  $\omega_c=eB/m$, $m$ is effective mass, $k_B$ is the Boltzmann constant, $\Delta$ is the broadening of the Landau levels (LL).  An example of such a fitting procedure for $V_g=-5$~V within the  magnetic field range $(1.2-2.5)$~T is shown in Fig.~\ref{F3}(a) by the solid line.

To find the carrier density from the oscillations frequency $f$, one should know the spin and valley degeneracy, $K=s+v$, of the Landau levels:
\begin{equation}
p_{SdH}=f\frac{e}{h}\times K.
\label{eq03}
\end{equation}

\begin{figure}
\includegraphics[width=\linewidth,clip=true]{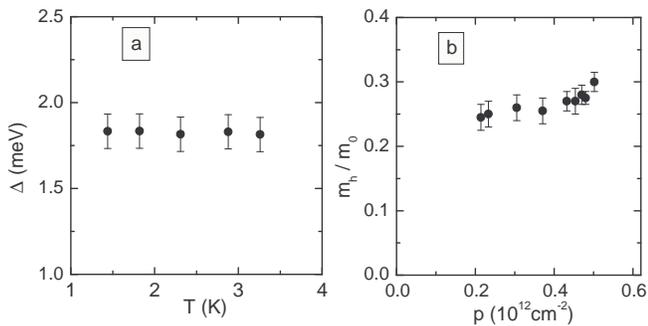}
\caption{(Color online) (a) -- The broadening $\Delta$ for different temperatures at $p=3.6\times 10^{11}$~cm$^{-2}$. (b) -- The values of the hole effective mass found at different hole densities.}
\label{F5}
\end{figure}

The most straightforward way to determine $K$ is based  on analysis of the positions of the minima  of $\rho_{xx}$ in {reciprocal magnetic field $1/B_{min}$ plotted as function of the oscillation number $N$ [see Fig.~\ref{F3}(c)]. Really,  the minima of $\rho_{xx}$ occur in those fields, when the integer numbers of Landau levels are filled. The number of states in nondegenerate LL is $(e/h)B$  therefore at carrier density $p$,  the filling factor will be equal to $\nu=p/\left[(e/h)B\right]$, so for integer $\nu$ [that corresponds to the minima in $\rho_{xx}(B)$]  $1/B_{min}(N)= (e/h)K\times N/p$
\footnote{It should be noted that the  $1/B_{min}$ versus $N$ dependence extrapolates to zero at $N=0$. It does not mean that the Berry phase is equal to zero. In 2D systems in which carrier density, but not the Fermi energy,  is constant with the growing magnetic field, the $1/B_{min}$ versus $N$ dependence extrapolates to zero at $N=0$ regardless of the value of the Berry phase.}.

Above, we have shown that the hole density is equal to $p_H$, therefore we have plotted in Fig.~\ref{F3}(c) the dependences $1/B_\text{min}(N)=(e/h)\times KN/p_H$ for different values of $K$. It is evident that the experimental values of  $1/B_\text{min}(N)$ coincide with $(e/h)\times KN/p_H$ with $K=2$ only.
The hole density  found from the frequency of the SdH oscillations using the degeneracy $K=2$    are shown for all $V_g$ values in Fig.~\ref{F2}(b). One can see that it coincides with $Q/e$ determined from the capacitance measurements  and with $p_H$ over the whole density range.
So, all the above data and their analysis show that the states of the top of valence band at hole density less than $5\times10^{11}$~cm$^{-2}$ are two-fold degenerate.

\begin{figure}
\includegraphics[width=\linewidth,clip=true]{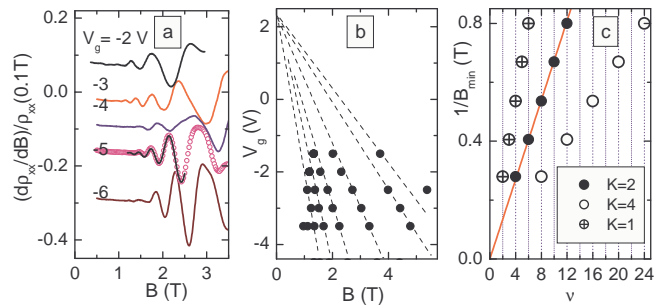}
\caption{(Color online) (a) -- The magnetic field dependencies of $(d\rho_{xx}/dB)/\rho_{xx}(0.1\text{ T})$ at different $V_g$, $T=1.4$~K. The solid line for $V_g=-5$~V is fit to the Lifshits-Kosevich formula, Eq.~(\ref{eq01}). (b) -- The positions of the maxima of  $\rho_{xx}$ in coordinates $(B,V_g)$. The dashed lines are provided as a guide for the eye. (c) -- The points are the positions of minima of $\rho_{xx}$ in inverse magnetic field, $1/B_\text{min}$, as a function of the  filling factor for the  different degeneracy of LL. The solid line is $(e/h)\times (N/p_H)$. }
\label{F3}
\end{figure}

\begin{figure}
\includegraphics[width=\linewidth,clip=true]{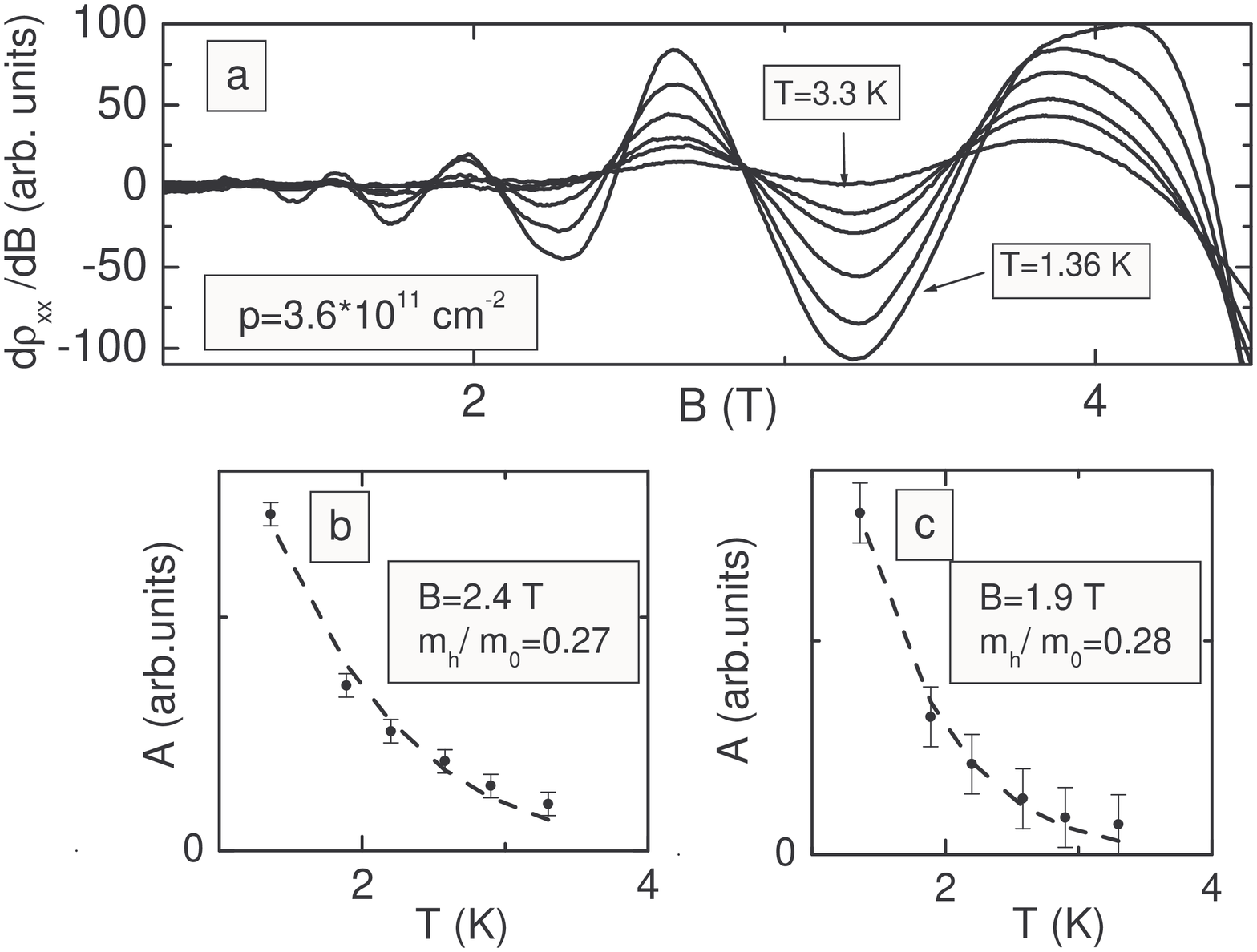}
\caption{(Color online) (a) -- the magnetic field dependences of  $d\rho_{xx}/dB$ taken at different temperatures at $p=3.6\times 10^{11}$~cm$^{-2}$. (b) and (c) -- The amplitude of the oscillations at different temperatures for $B=2.4$~T and $1.9$~T, respectively (symbols). The dashed curves are the results of the best fit to Eq.~(\ref{eq02}) with $m_h/m_0$ as the  fitting parameter. }
\label{F4}
\end{figure}

Another important parameter of the spectrum is the hole effective mass $m_h$.  The value of $m_h$ can be found from the temperature dependence of the amplitude of SdH oscillations. As an example, the oscillations of  $d\rho_{xx}/dB$ at hole density $3.6\times10^{11}$~cm$^{-2}$ measured at different temperatures are presented in Fig.~\ref{F4}(a). To exclude the monotone contribution, analysis of temperature behavior of $d\rho_{xx}(B)/dB$ instead of $\rho_{xx}(B)$} is more reliable for such studies. The temperature dependencies of the oscillations amplitude  at two magnetic fields are shown in Figs.~\ref{F4}(b) and ~\ref{F4}(c). They are well described by  Eq.~(\ref{eq02}) with the hole effective mass $m_h=(0.27-0.28)\,m_0$. It should be noted that the use of Eq.~(\ref{eq02}) for determination of the effective mass presumes that the broadening $\Delta$ is independent of temperature. To check the validity of this assumption we have determined  $\Delta$ at different temperatures from the magnetic field dependencies of the oscillations amplitude.  Figure~\ref{F5}(a) demonstrates that  $\Delta$ really is independent of $T$ in our case.

The values of the hole effective mass were determined for all the structures investigated with $d$ from $8.3$~nm to $20.4$~nm at different hole densities. These results are collected  in Fig.~\ref{F6}. It is evident that: (i) the hole effective mass is independent of the quantum well width within experimental accuracy, (ii)  $m_h=(0.22\pm 0.03)\,m_0$ at hole density $p\simeq (1-2)\times 10^{11}$ ~cm$^{-2}$, (iii) it weakly increases with the hole density.

\section{The calculations of the energy spectrum of the valence band and
discussion}
\label{sec:Calculation and discussion}

This section is devoted to theoretical consideration of the electron spectrum in HgTe quantum well.

\subsection{The energy spectrum of the valence band of the well with
[013] orientation}

\begin{figure}[b]
\includegraphics[width=0.8\linewidth,clip=true]{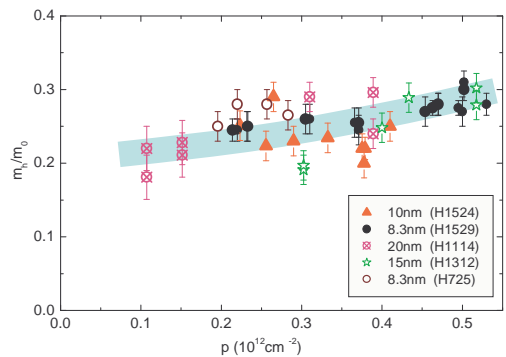}
\caption{(Color online) The values of hole effective mass measured in all the structures investigated within whole hole density range. The calculated dependence of $m_h$  with different parameters is presented by the wide semitransparent line (see Section~\ref{ssec3c}).}
\label{F6}
\end{figure}

The band structure was calculated using a 4-band Kane Hamiltonian. For
the quantum well grown on the plane (013)  the explicit form of the
Hamiltonian including deformation terms is given in \cite{Zholudev12}. Calculation of
the wave functions and electron spectrum was carried out for the
superlattice with a large period when barriers virtually nontransparent.
In this method, one can use periodic boundary conditions
for the electron wave function (corresponding to the center of the
Brillouin zone in the superlattice) and expand the wave function in a
series:
\begin{eqnarray}
\psi_j(x,y,z)&=&\exp\left(ik_xx+ik_yy\right)\nonumber \\
&\times&\sum_{l=-N_{max}}^{l=N_{max}}C_l(k_x,k_y)\exp\left(\frac{2i\pi
zl}{d_{sl}}\right)\nonumber \\
&=&\exp\left(ik_xx+ik_yy\right)\varphi_j(k_x,k_y,z)
\end{eqnarray}
where the index $j$ denotes the component of the wave function,
$k_x$ and $k_y$ are components of the electron wave vector in the plane of the
quantum well ($x$ and $y$ axes are chosen along the $[100]$ and $[03\bar
1]$ directions, respectively), $d_{sl}$ is the  superlattice period, $C_l(k_x,k_y)$ are the expansion
coefficients. With this approach, finding the electron spectrum and the
wave functions reduces to finding the eigenvalues and eigenvectors of
the matrix $8 (2N_{max} + 1)\times 8 (2N_{max} + 1)$. In the calculation, the value $N_{max}$ = 20 was used. Check with $N_{max}$ = 30 showed
practically identical results. We used in calculations the parameters
for Hg$_{1-x}$Cd$_x$Te from  Ref.~\cite{Novik05}. Parameters to describe the
deformation contribution to the Hamiltonian were taken from Ref.~\cite{Takita}.
\begin{figure}
\includegraphics[width=0.9\linewidth,clip=true]{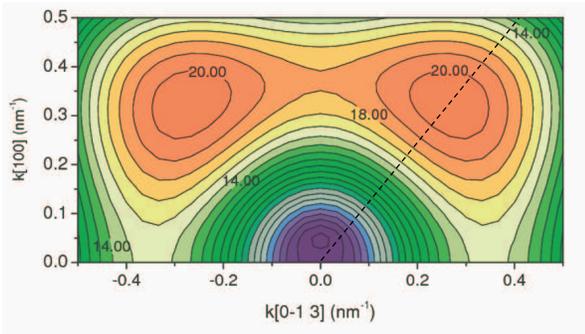}
\caption{(Color online) The isoenergy contours of the top of valence band. Orientation of substrate is (013), width of the well is equal to $8.3$~nm. We show half of the picture, the second part is a mirror image. The energy is measured from $E(k=0)$. The step is $1$~meV.  The energies for  some contours are shown in plot. The dashed line shows the direction which passes through the  maximum of $E(\mathbf{k})$.}
\label{F7}
\end{figure}

\begin{figure}[b]
\includegraphics[width=\linewidth,clip=true]{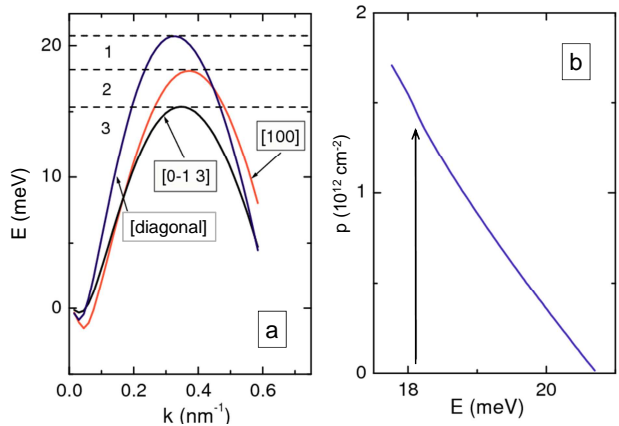}
\caption{(Color online) (a) -- The dependencies $E(k)$ for three directions: $[100]$, $[0\bar{1}3]$  and \emph{[diagonal]} one, which passes through the  maximum of $E(\textbf{k})$. (b) -- The dependence of hole density on the Fermi energy. The arrow shows the density of holes at which the  two nearby valleys are merged. }
\label{F8}
\end{figure}

Let us consider the calculated spectrum of the valence band in the
quantum well with $d=8.3$~nm for which the experimental
results were presented in Figs.~\ref{F2}--\ref{F6}. Figure~\ref{F7} shows that the top of the
valence band consists of 4 local extremes (valleys), which are located
at $k\neq 0$. The $E$~versus~$k$ dependencies for three directions [100],
$[03\bar 1]$ and direction  which passes through  maximum (labeled as \emph{[diagonal]})  are presented in Fig.~\ref{F8}(a). The dashed lines in this figure
divide the energy into three regions: (i) within the region 1, there are
4 closed isoenergy contours. It means that  the states are 8-fold degenerate  with taking into account the spin degeneracy; (ii) within the region 2,
there are 2 closed isoenergy contours, which arise  after the merger of
the two pairs nearest valleys and the states are 4-fold degenerated;
(iii)  within the region 3, there are 2 concentric contours.

The dependence of the hole density on the Fermi energy for $p<2\times 10^{12}$~cm$^{-2}$  is
plotted in Fig.~\ref{F8}(b). The hole density  $p$ was calculated as
 $p=svS_1(E_F)/(2\pi)^2$, where $S_1(E_F)$ is the area of the single
closed contour at energy $E_F$ in $k$ space. For this case $s=2$ while $v=4$ within the Fermi energy range $(20.5-18.0)$~meV and $v=2$ for lower energy when two valleys are merged.   This figure shows that over whole studied density range
($p<5\times 10^{11}$~cm$^{-2}$) there are 4 isolated isoenergy contours.

At this point it should be noted that all the information on the
spectrum was obtained from the oscillations in the magnetic field.
Therefore, for comparison of the theoretical calculations with experiment
one needs to know the energies of the Landau levels $E_N(B)$ for such
spectrum.  We use quasiclassical quantization rule which for
quantization by a magnetic field is reduced to replacement $S_1\rightarrow
2\pi(N+1/2)/L_H^2$, where $L_H=\sqrt{\hbar /eB}$ is the magnetic length, $N=0,1,2,...$. So,
the spin and valley degeneracy of  the Landau levels coincides with the
degeneracy of the spectrum at $B=0$ in the valence band top. The energy
distance between nearest Landau levels ($\hbar\omega_c$) is equal to $\hbar\omega_c=
2\pi(eB/\hbar )(dS_1/dE)^{-1}$ and the effective mass is
$m=(\hbar^2/2\pi)dS_1/dE$. Namely this value is determined from the
temperature dependence of the amplitude of the SdH oscillations [see
Figs.~\ref{F4}(b) and \ref{F4}(c)]
\footnote{Note that all the calculations of the Landau levels, performed until now
(see, e.g., Ref.~\cite{Koenig08, Novik05}), inadequately describe the behavior of the Landau
levels of the valence band in weak magnetic fields for quantum wells
with $d>(7\sim 8)$~nm. It is easy to understand if we take into account the
following considerations. The valence band top in these quantum wells is
realized in the side valleys where the effective electron mass is
negative. Therefore, in weak magnetic fields Landau levels should move
down with increasing magnetic field. However, the calculations presented
in the literature \cite{Koenig08, Novik05} show the movement of the Landau levels upwards.}.

Thus, the theoretical calculation predicts 8-fold degeneracy of the
valence band states while experimental data give 2-fold degeneracy
\footnote{Such discrepancy forced the authors of \cite{Minkov13} to use the isotropic
approximation for  analysis of the experimental data. At this approximation
the valleys are absent and isoenergy contours are concentric circles.
However, in this case, the values of the hole effective mass differ
significantly from theoretical ones.}.

{\bf What factors were not taken into account in the above calculation?}
First and foremost is the lack of inversion symmetry in the
heterostructures based on $A_3B_5$, $A_2B_6$ semiconductors. It was
shown \cite{Tarasenko15,Dai,Minkov16} that interface inversion
asymmetry (IIA) significantly changes the energy spectrum of HgTe
quantum well,  especially the top of the valence band.

\subsection{The energy spectrum of the valence band with taking into
account IIA}
To take into account the IIA we used an additional term in the
Hamiltonian, which is suggested by E.~Ivchenko \cite{Ivchenko05}. This term leads to ``the mixing of the states'' at the boundaries and in case
when the normal to the quantum well is along to the [013] direction, is as follows
\begin{eqnarray}
H_i&=&\frac{dU}{dz}\frac{1}{\sqrt 3}\times \label{eq04} \\
&&\left(
   \begin{array}{cccccccc}
0 & 0 & 0 & 0 & 0 & 0 & 0 & 0 \\
0 & 0 & 0 & 0 & 0 & 0 & 0 & 0 \\
0 & 0 & 0 & \frac{3}{5} &-i\frac{4}{5} & 0 & -\frac{3\sqrt{2}}{10} &
i\frac{4\sqrt{2}}{5} \\
0 & 0 & \frac{3}{5} & 0 & 0 & -i\frac{4}{5} & 0 & \frac{3\sqrt{6}}{10}\\
0 & 0 & i\frac{4}{5} & 0 & 0 & -\frac{3}{5} & \frac{3\sqrt{6}}{10} & 0\\
0 & 0 & 0 & i\frac{4}{5} &  -\frac{3}{5} & 0 & i\frac{4\sqrt{2}}{5} &
-\frac{3\sqrt{2}}{10}\\
0 & 0 & -\frac{3\sqrt{2}}{10} & 0 & \frac{3\sqrt{6}}{10} &
-i\frac{4\sqrt{2}}{5} &
0 & 0 \\
0 & 0 & -i\frac{4\sqrt{2}}{5} & \frac{3\sqrt{6}}{10} & 0 &
-\frac{3\sqrt{2}}{10} & 0 & 0\\
\end{array}
\right)\nonumber
\end{eqnarray}
where $z$ is the normal to the quantum well, the function $U (z)$ depends
only on jump in the semiconductor composition at the boundary. We have
assumed it is a linear function of the Cd content $x$
\begin{equation}
U(z)=[1-x(z)]g_4.
\label{eq05}
\end{equation}
and $g_4$ is parameter of ``mixing''.

\begin{figure}
\includegraphics[width=\linewidth,clip=true]{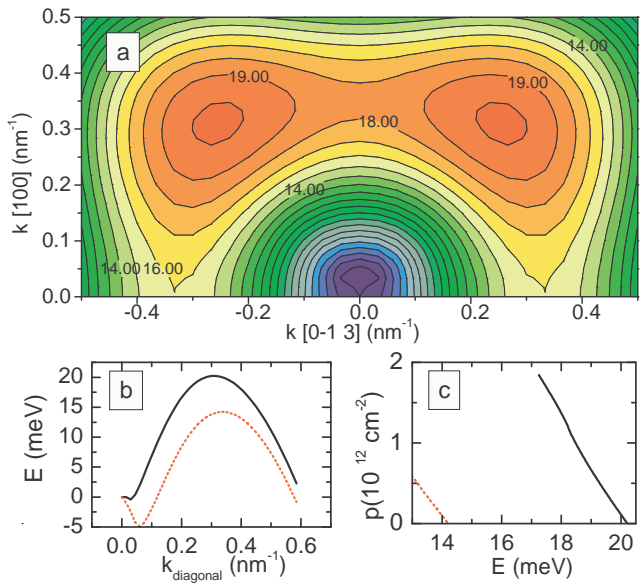}
\caption{(Color online) (a) -- The isoenergy contours of the upper split-off state  calculated with taking into account IIA. (b) -- The $E$~versus~$k_\text{diagonal}$ dependences for both split-off states.  (c) -- The dependence of hole density in the upper (solid curve) and lower (dashed curve)  split-off states on the Fermi energy.}
\label{F9}
\end{figure}

The results of such calculation are presented in Fig.~\ref{F9}.  We used the
value $g_4=0.8$~eV\AA\,  (the value $g_4=0.8-0.6$~eV\AA\, satisfactorily describes the
experimental data on the splitting of the valence and conduction bands
in the structures with $d \sim d_c$ \cite{Minkov16}). Figure~\ref{F9}(b) shows that IIA leads
to large (about $(6-7)$~meV) ``spin'' splitting of the states in the valleys while
the isoenergy contours of the upper states remain close to that
calculated without IIA [compare Fig.~\ref{F7} and Fig.~\ref{F9}(a)]. As Figs.~\ref{F9}(b) and~\ref{F9}(b) show,
the splitting is so large that up to hole density of $\sim 4\times
10^{12}$~cm$^{-2}$ the holes fill the upper split  states only.
Thus, taking into account the  interface inversion asymmetry reduces the degeneracy of
the top of valence band from 8 to 4 within studied hole density range.
But this is not enough, because the experimental value of the degeneracy
is 2.

What else dos not take into account this theoretical model?

\subsection{The role of the additional asymmetry of the quantum well resulting from difference of the mixing at the well boundaries and their broadening }
\label{ssec3c}

In the previous section, we considered the role of mixing the states on the boundaries of a quantum well in the simplest symmetric case, when the additional term, Eq.~(\ref{eq04}), to the Hamiltonian is proportional to $dU/dz$ and differs on the boundaries only by sign due to $x(z$) [see Eq.~(\ref{eq05})]. In this case, IIA removes the spin degeneracy, but leaves a valley degeneracy. As further calculations of the spectrum show, this state is unstable: the additional asymmetry of the boundary conditions, i.e. different values of the parameter $g_4$ at the boundaries,  the different broadening of the walls of the quantum well  lead to the removal of valley degeneracy, too.  The reasons for this asymmetry relate to the fact that the walls of the quantum well are grown under different conditions (at different times and temperature). In addition, the electric field, normal to the 2D plane, both built-in, which results from uncontrolled doping of barriers, and created by a gate voltage, also leads to the removal of valley degeneracy.

\begin{figure}
\includegraphics[width=\linewidth,clip=true]{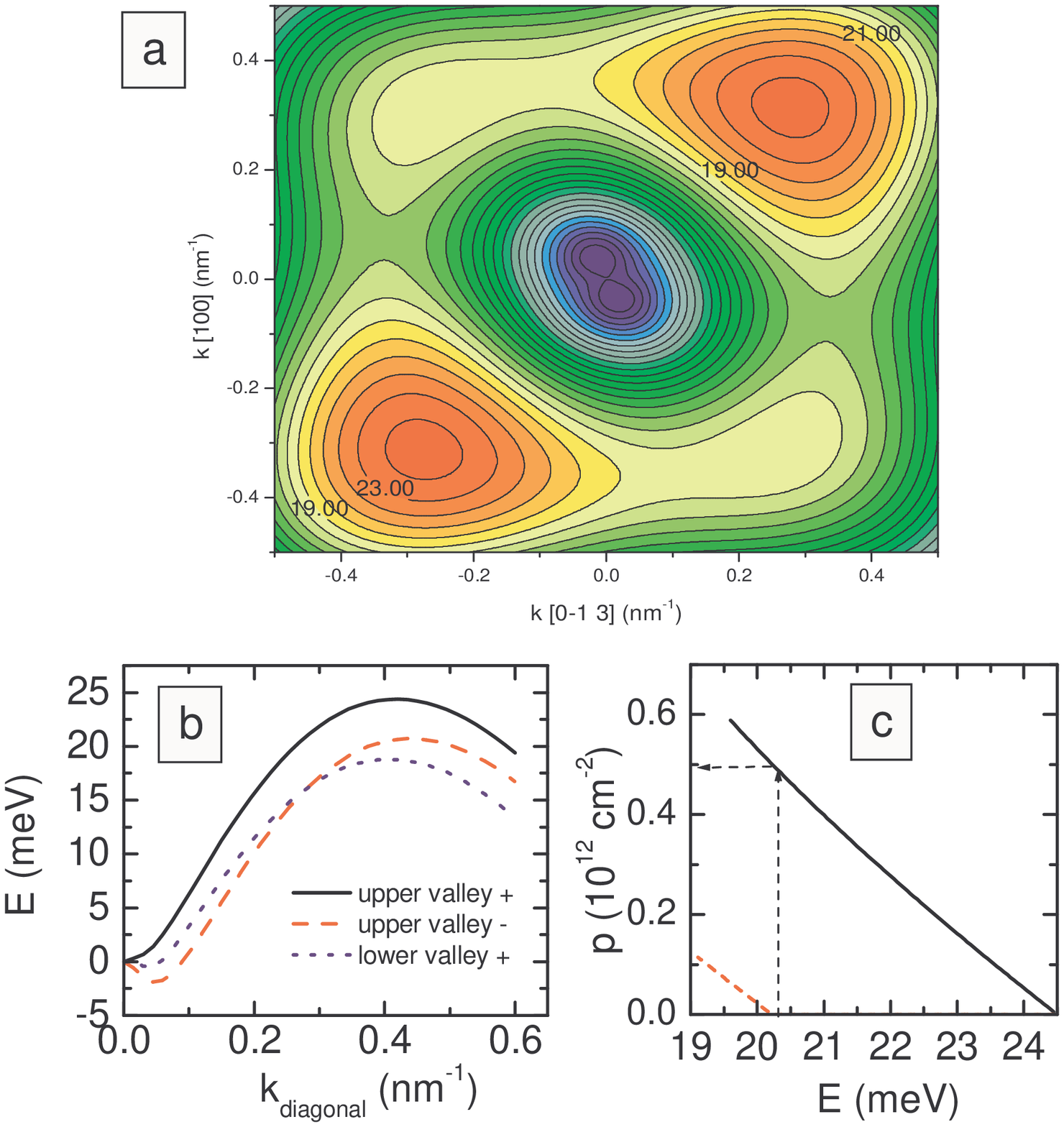}
\caption{(Color online) (a) -- The isoenergy contours of the upper split-off states calculated with the different  parameter $g_4$, namely $g_4=1$~~eV\AA\, and $g_4=0.8$~~eV\AA\, on one and another boundaries. (b) -- The $E$~versus~$k_\text{diagonal}$ dependences for both split-off states of upper valley (solid and dashed lines) and for upper states of lower valley (the dotted line). (c) -- The dependence of hole density in the upper (solid curve) and lower (dashed  curve)  split-off states on the Fermi energy.}
\label{F10}
\end{figure}

As an example, in Fig.~\ref{F10} we plot the results of the calculation of the spectrum with an insignificant  difference in the parameter $g_4=1$~eV\AA\, and $g_4=0.8$~eV\AA\, for one and another boundaries (in Fig.~\ref{F9}, the value of parameter $g_4$ was equal to $0.8$~eV\AA\, for both boundaries). A comparison of the isoenergy contours in Fig.~\ref{F9}(a) and  Fig.~\ref{F10}(a)  shows that such difference removes the valley degeneracy.
As seen from Fig.~\ref{F10}(b)  the ``spin'' splitting of the valleys practically
unchanged, however the height of the valleys becomes different. This
figure demonstrates, that within the Fermi energy range $(24.5-20.5)$~meV,
the holes fill only two from four ``valleys''.  This means that in this
range, the degeneracy became 2 rather than 4.
The dependence of the hole density on the Fermi energy presented in Fig.~\ref{F10}(c)
shows that the degeneracy 2  should be up to the hole density of about $5\times
10^{11}$~cm$^{-2}$. The arrow in this picture marks the energy when lower
``valleys'' begin to fill up that leads to increase of the rate $dp/dE_F$.

Thus, taking into account both the interface inversion asymmetry and
difference in parameters of the wall of the quantum well  allows us to
explain the degeneracy of the top of valence band states \footnote{It should be noted that the degeneracy two for our case is valley degeneracy of two single-spin valleys. However  ``spin'' states of these valleys are different therefore the magnetic field should split LL due to the Zeeman effect. We assume  the distortion of the SdH oscillations at $B>2.5$~T evident from Fig.~\ref{F3}(a) is  result  of such splitting. At $B<2.5$~T the Zeeman splitting is small and does not lead to splitting of the SdH oscillations.}

\begin{figure}
\includegraphics[width=0.9\linewidth,clip=true]{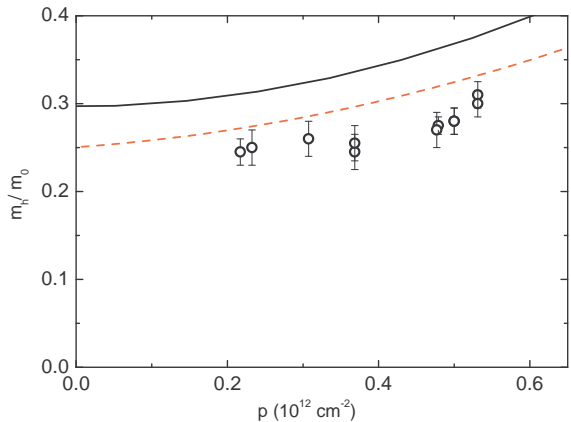}
\caption{(Color online) The experimental (symbols) and theoretical (lines) hole density dependence of the hole effective mass in the structure with $d=8.3$~nm.  The solid line is calculated when the value of the parameter $g_4$ is the same for both boundaries of the quantum well, $g_4=0.8$~~eV\AA\,.   The dashed line is the case when the value of $g_4$ is different for different boundaries, $g_4=1$~eV\AA\, and $0.8$~eV\AA\,.}
\label{F11_1}
\end{figure}

Now consider the dependence of the holes effective mass on their density. The calculated dependences for the well with $d=8.3$~nm  for the two cases considered above are shown in Fig.~\ref{F11_1}. The solid  curve shows the $m_h$~versus~$p$ dependence  calculated for the case when the parameter $g_4$ for both boundaries was equal to 0.8~eV\AA\,. The dashed curve is the case when $g_4=0.8$~eV\AA\, and $1$~eV\AA\, for different boundaries. It can be seen that the values and the dependences on the density are quite close to each other. In this figure we present the experimental $m_h$ values  for the sample H1529 with $d=8.3$~nm. They agree  well with the calculated dependence for the second case.

Calculations carried out for wider wells, $d=(10-20)$~nm, show that taking into account only the difference in the parameter $g_4$ on the boundaries is not sufficient to remove the valley and spin degeneracy in the hole density range up to  $p=(5-6)\times 10^{11}$~cm$^{-2}$ for which experimental results were presented in Fig.~\ref{F6}.

However, the calculations carried out within some modifications of the theoretical model which take into account other causes of asymmetry, the difference in broadening of the boundaries of the well, the electric field, normal to the 2D plane, show that it is possible to obtain the sufficient values of the spin and valley splitting. This is achieved with reasonable values for the parameters: the boundaries broadening is always less than $0.1d$, the electric field is less than $\simeq6\times 10^5$~V/cm, and $g_4=(0.6-1.2)$~eV\AA\,.
Therewith, the range of change of the effective mass, which is shown by the wide semitransparent line in Fig.~\ref{F6}, is not wide and
well describes the data obtained in all the structures investigated.  It is currently not possible to separate the roles of various mechanisms.

\section{Conclusion}
We presented the experimental results of the studies of  the longitudinal and transverse magnetoresistance in HgTe/Cd$_x$Hg$_{1-x}$Te quantum wells of $(8.3-20.4)$~nm nominal width in the hole conductivity regime in a wide range of hole density. Comparison of the Hall density with the density found from the period of the SdH oscillations clearly shows that the degeneracy of states  of the top of the valence band is equal to 2 over the entire hole density range from $1\times 10^{11}$~cm$^{-2}$ to $5.5\times 10^{11}$~cm$^{-2}$. This value of the degeneracy does not agree with the calculations of the spectrum performed within the framework of the $4$-bands $kP$-method for symmetric quantum wells. These calculations show that the top of the valence band consists of four spin-degenerate  extremes located at $k\neq 0$, which gives the total degeneracy $K=8$. It is shown that taking into account the mixing of states at the interfaces \cite{Ivchenko05} leads to the removal of the spin degeneracy, which reduces the degeneracy to $K=4$. However, such a state is unstable. Taking into account any additional asymmetry (the difference in the mixing parameters at the interfaces, the different broadening of the boundaries of the well, etc.) leads to the removal of valleys degeneracy, making $K=2$ within the investigated range of hole density.
It is noteworthy that for our case two-fold degeneracy occurs due to degeneracy of two single-spin valleys.

Another parameter of the energy spectrum, the hole effective mass, was experimentally determined in the entire range of the hole density by analyzing  the temperature dependence of the amplitude of the SdH oscillations. We have shown that the effective mass at  $p\simeq 2\times 10^{11}$~cm$^{-2}$ in the wells of width $d=(8-20)$~nm is equal to $(0.25\pm0.02)\,m_0$ and weakly increases with the increasing hole density. Such a value of $m_h$ and its dependence on  $p$ are in a good agreement with the calculated effective mass of the valleys.

\acknowledgements

We are grateful to E.L. Ivchenko and  M.~Nestoklon  for useful discussions.
The work has been supported in part by the Russian Foundation for Basic
Research (Grants No. 16-02-00516, No. 15-51-06001 and No. 15-02-02072),  by  Act 211 Government of the Russian Federation, agreement No.~02.A03.21.0006, and  by  the Ministry of Education and Science of the Russian
Federation under Project No.~3.9534.2017/BP.

%

\end{document}